\documentclass[superscriptaddress,twocolumn,floatfix]{revtex4-2}
\usepackage{graphicx}
\usepackage{natbib}
\usepackage{amsmath,amsthm,amssymb,dsfont,tabularx,soul}
\usepackage{mdframed}
\usepackage{hyperref}
\newcommand{\ket}[1]{\left| #1 \right\rangle}
\newcommand{\bra}[1]{\left\langle #1 \right|}
\newcommand{\proj}[1]{\ket{#1}\hskip-1mm\bra{#1}}

\newcommand{\braket}[2]{\langle #1|#2 \rangle}
\newcommand{\ketbra}[2]{\left|#1\right\rangle\hskip-1mm\left\langle#2\right|}

\newtheorem{claim}{Claim}
\newtheorem{inference}{Inference}

\begin{document}

\title{Contextuality, Coherences, and Quantum Cheshire Cats}
\author{Jonte R. Hance}
\email{jonte.hance@bristol.ac.uk}
\affiliation{Department of Quantum Matter, Graduate School of Advanced Science and Engineering, Hiroshima University, Kagamiyama 1-3-1, Higashi Hiroshima 739-8530, Japan}
\affiliation{Quantum Engineering Technology Laboratories, Department of Electrical and Electronic Engineering, University of Bristol, Woodland Road, Bristol, BS8 1US, UK}
\author{Ming Ji}
\affiliation{Department of Quantum Matter, Graduate School of Advanced Science and Engineering, Hiroshima University, Kagamiyama 1-3-1, Higashi Hiroshima 739-8530, Japan}
\author{Holger F. Hofmann}
\email{hofmann@hiroshima-u.ac.jp}
\affiliation{Department of Quantum Matter, Graduate School of Advanced Science and Engineering, Hiroshima University, Kagamiyama 1-3-1, Higashi Hiroshima 739-8530, Japan}

\begin{abstract}
We analyse the quantum Cheshire cat using contextuality theory, to see if this can tell us anything about how best to interpret this paradox. We show that this scenario can be analysed using the relation between three different measurements, which seem to result in a logical contradiction. We discuss how this contextual behaviour links to weak values, and coherences between prohibited states. Rather than showing a property of the particle is disembodied, the quantum Cheshire cat instead demonstrates the effects of these coherences, which are typically found in pre- and postselected systems.
\end{abstract}

\maketitle

\section{Introduction}
The quantum Cheshire cat protocol has raised many eyebrows. The protocol preselects and postselects states where the weak value for the spatial projection operator of a quantum particle (e.g. a photon) is zero along a given path. This is despite this pre- and postselection giving a non-zero weak value for an operator supposedly representing one of the particle's constituent properties along that path (e.g. its polarisation). This is often interpreted as the property becoming disembodied, travelling along a path the particle itself cannot traverse \cite{jayaraman2020sensationalism}. The protocol was initially given as a thought experiment \cite{Aharonov2013Cheshire}, but later demonstrated experimentally \cite{Denkmayr2014QCCExp,Ashby2016QCCExperiment,sponar2016fundamental,sponar2018weak,nawaz2019atomic,Danner2023Neutrons}. Recent work claims to have extended the protocol to dynamically changing the disembodied property \cite{Aharonov2021Dynamical,Hance2022DQCCDetectable}, swapping this disembodied property between two particles \cite{Das2020QCCSwitch,Liu2020QCCExchangeExp,Danner2023Neutrons}, delayed choice of which path carries the particle and which carries the disembodied property \cite{Das2021Delayed,Wagner2023Delayed}, disembodying multiple properties simultaneously \cite{pan2020multi,danner2023three}, and even ``separating the wave-particle duality'' of a particle \cite{li2023experimental}.

However, this interpretation of the protocol is controversial. Many have questioned how paradoxical the protocol actually is \cite{Sokolovski2016}, with some saying the effect simply constitutes standard quantum interference \cite{Correa2015QCCInterference} or entanglement {\cite{saeed2021quantum}}{\cite{Rameez2019Entanglement}}, and others claiming the same result can be obtained using classical physics \cite{Atherton2015QCCClassical}.

In this paper, we analyse the protocol using the recently-developed tool of contextuality theory \cite{Budroni2022ContextualityReview}. This involves considering the relationships between different possible (ideally local) measurements, and our classical intuitions about what those measurements would infer, and then seeing whether those inferences are compatible in the scenario in question. More formally, it involves assigning a measurement context to every mutually-orthogonal set of measurement outcomes, observing that some outcomes are shared by multiple contexts, and using these shared outcomes to infer how these contexts relate to one another.

While \cite{yu2014quantum} and \cite{waegell2018contextuality} mention in passing that the quantum Cheshire cat scenario is equivalent to a contextual scenario known as the 2-qubit Peres-Mermin square, this has not yet actually been shown, nor has the scenario been fully analysed using contextuality theory. We seek to correct this oversight.

To do so, we identify a set of observable properties, that allow us to derive statements about a quantum particle's path and polarisation from a preselected initial state and a postselected final state. We then show that these observable properties belong to different measurement contexts, where the postselected result should be impossible because it implies a contradiction between the polarisation, the path, and the correlation between the two. By analysing the Hilbert space algebra, we then proceed to show that the contextuality argument links naturally to the Cheshire cat argument presented in \cite{Aharonov2013Cheshire}. The original Cheshire cat argument emphasised the contradiction between the pairing of correlation and polarisation on the one hand, and path on the other, showing that the path determined by polarisation and correlation was opposite to the independently-determined path. We find that contextuality emphasises the symmetry of the three statements regarding path, polarisation, and correlation. However, other pairings are possible to highlight the contradictions. The impression that the quantum Cheshire cat describes the disembodiment of a physical property of the particle is therefore a consequence of a very specific interpretation of the contextuality relations that characterise the paradox.

This paper is organised as follows. In Section \ref{sect:Prot}, we go over the quantum Cheshire cat protocol, with a couple of simplifying adaptations.
In Section \ref{sect:Context}, we show we can define three claims about the properties of a particle in the quantum Cheshire cat protocol: individually, each claim can be shown experimentally to be true, but combining the three Claims leads to a contradiction. We then go on to adapt this logic to form an inequality, which our classical intuition would expect to be valid, but which a quantum mechanical description of the quantum Cheshire cat protocol violates.
In Section \ref{sect:Coherences} we use weak values and coherences to decompose the statistical operator describing the quantum Cheshire cat scenario into different bases. In each of these bases, we see coherences between modes which are not occupied. We show that these coherences between prohibited states causes the contextual behaviour demonstrated by the protocol.
In Section \ref{sect:Meaning}, we discuss the meaning of these coherences between prohibited states, and how they link to the compound operators in the original Quantum Cheshire Cats paper. We show that the meaning of these coherences becomes more obvious when we combine Claims differently. We then summarise our findings in Section \ref{sect:Conc}.

\section{The Cheshire Cat Protocol}\label{sect:Prot}

In this Section, we describe a quantum Cheshire cat protocol, where, by choosing a suitable pre- and postselection, a quantum particle appears to become separated from one of its properties. This protocol is of the same form as the one introduced in \cite{Aharonov2013Cheshire}, however, our form is both simplified, and emphasises certain key features. We implement this protocol optically---the quantum particle is a photon, and the ``disembodied'' property its polarisation.

\begin{figure}[ht]
    \centering
    \includegraphics[width=\linewidth]{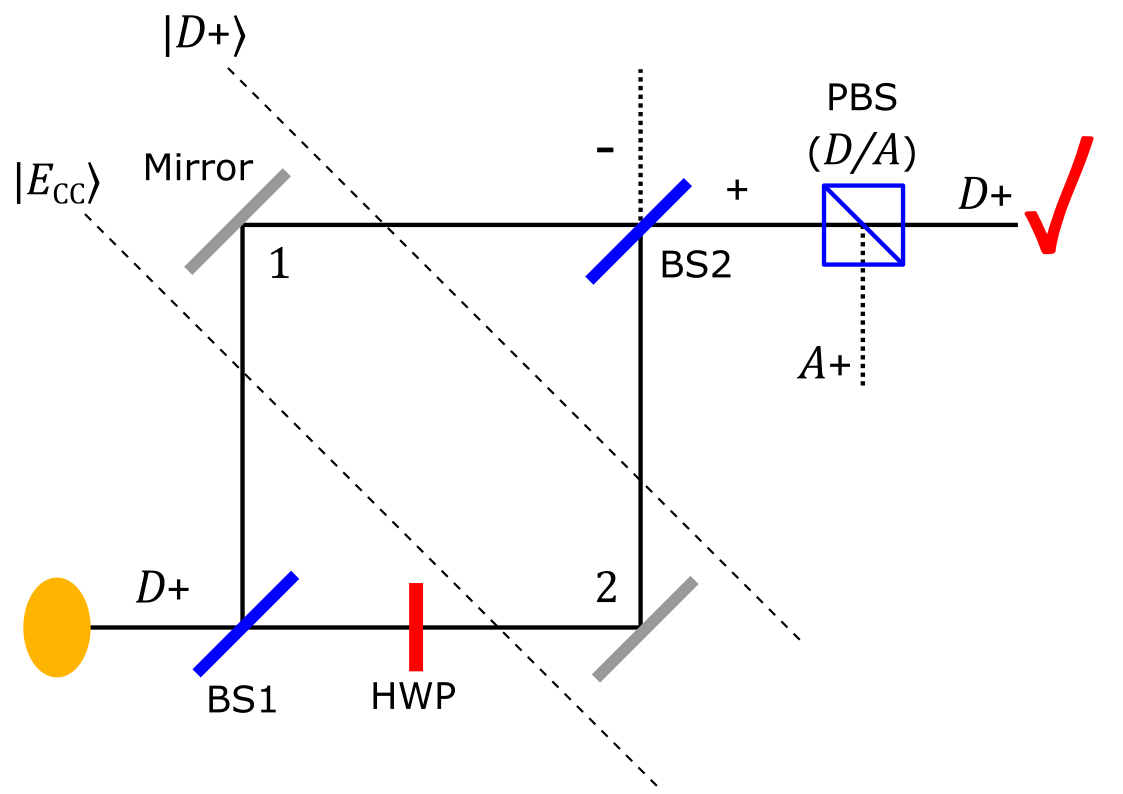}
    \caption{Optical implementation of the quantum Cheshire cat protocol. BS1 and BS2 are balanced beamsplitters, PBS ($D/A$) is a polarising beamsplitter which reflects anti-diagonally ($A$) polarised light and transmits diagonally ($D$) polarised light. HWP is a half-wave plate, aligned to transform $D$-polarised light into $A$-polarised light (and vice-versa). This implementation preselects the photon in entangled state $\ket{E_\text{CC}}$, and postselects it in state $\ket{D+}$. This creates a quantum Cheshire cat---the photon passes along arm 1, but its polarisation appears to travel along arm 2. See Section \ref{sect:Prot} for details.}
    \label{fig:HolgerQCC}
\end{figure}

In the quantum Cheshire cat protocol (as given in Fig.~\ref{fig:HolgerQCC}), a diagonally ($D$) polarised single-photon is emitted from a source, and passed through a 50:50 beamsplitter (BS1). This puts the photon onto a superposition of paths 1 and 2 through the interferometer---specifically, the superposition $\ket{+}$, where we define superpositions $\ket{+}$ and $\ket{-}$ as
\begin{equation}
    \ket{\pm} = \frac{1}{\sqrt{2}}\left(\ket{1}\pm\ket{2}\right)
\end{equation}

Path 2 then passes through a half wave plate (HWP), aligned to cause a phase shift of $\pi$ between horizontal ($H$) and vertical ($V$) polarised components, such that it flips diagonal to anti-diagonal ($D$ to $A$) polarisation, and vice-versa. This preselects the photon in the entangled state $\ket{E_\text{CC}}$. In the $H/V$ basis, this can be represented as
\begin{equation}
\begin{split}
    \ket{E_\text{CC}} &= \frac{1}{2}\left(\ket{H1}+\ket{H2}+\ket{V1}-\ket{V2}\right)
\end{split}
\end{equation}
where for convenience we write
\begin{equation}
    \ket{ab} = \ket{a}\otimes\ket{b}
\end{equation}
and $H$ and $V$ polarisation are related to $D$ and $A$ polarisation by
\begin{equation}
    \begin{split}
        \ket{H} &= \frac{1}{\sqrt{2}}\left(\ket{D}+\ket{A}\right),\\
        \ket{V} &= \frac{1}{\sqrt{2}}\left(\ket{D}-\ket{A}\right)
    \end{split}
\end{equation}

Paths 1 and 2 then recombine at another 50:50 beamsplitter (BS2), before passing through a polarising beam splitter (PBS). This PBS transmits $D$-polarised light, and reflects $A$-polarised light. We then postselect on the photon being in state
\begin{equation}
    \ket{D+} = \frac{1}{2}\left(\ket{H1}+\ket{H2}+\ket{V1}+\ket{V2}\right)
\end{equation}
by only considering cases when the photon ends up transmitted through this PBS.
For the preselected input state $\ket{E_\text{CC}}$, the probability of finding a photon in this postselected output is 1/4.

We now want to ask what the properties of this pre- and postselected photon are, in the path-and-$H/V$ basis. The standard way to observe the properties of a particle in quantum mechanics is through projective, or Von Neumann, measurements. However, we have already postselected the photon in an output that is an equal superposition of all basis states. Projective measurements on these basis states would change the state of the photon, meaning the results obtained in a sequential measurement wouldn't actually tell us about the undisturbed pre- and postselected system. Therefore, we need a way of observing the properties of the system without disturbing the system.

Weak values were originally claimed to be a way to infer information about an observable, between a fixed initial and final state without disturbing the evolution of the system \cite{Aharonov1988Weak,aharonov1990properties}. Despite originally being considered in the context of weak measurement, weak values have since been observed in settings other than those using weak measurement \cite{Cohen2018WeakStrong,wagner2023quantum}, and their meaning appears more subtle than initially thought \cite{Hofmann2011UncertaintyfromWeak,matzkin2019weak,Hofmann2020ContextfromWeak}.

The weak value $\langle \hat{O}\rangle_w$ of an operator $\hat{O}$ between preselection $\ket{i}$ and postselection $\ket{f}$ is given by
\begin{equation}
    \langle \hat{O}\rangle_w = \frac{\bra{f}\hat{O}\ket{i}}{\braket{f}{i}}
\end{equation}
Using the pre- and postselection given above, we can obtain weak values for properties of the particle between BS1 and the HWP.

The spatial projection operators, representing projection on path $1$ and path $2$ respectively, are
\begin{equation}
\begin{split}
    \hat{\Pi}(1) &= \mathds{1}\otimes\proj{1},\\
    \hat{\Pi}(2) &= \mathds{1}\otimes\proj{2}
\end{split}
\end{equation}
Their weak values are
\begin{equation}
\begin{split}
    \langle\hat{\Pi}(1)\rangle_w &= 1,\\
    \langle\hat{\Pi}(2)\rangle_w &= 0
\end{split}
\end{equation}
The original Cheshire cat paper \cite{Aharonov2013Cheshire} interprets these values as showing that a photon which passes the pre- and postselection must travel on path 1.

We can define the polarisation difference operator
\begin{equation}
\begin{split}
    \hat{\sigma}_{HV} &= \proj{H}-\proj{V}\\
    &= \ketbra{D}{A}+\ketbra{A}{D}
\end{split}
\end{equation}
From these projection and difference operators, we can define compound operators
\begin{equation}\label{eq:Compound}
    \begin{split}
        \hat{\sigma}_{HV}(1) &= \hat{\sigma}_{HV}\otimes\proj{1},\\
        \hat{\sigma}_{HV}(2) &= \hat{\sigma}_{HV}\otimes\proj{2}
    \end{split}
\end{equation}
The weak values of these compound operators are
\begin{equation}
    \begin{split}
        \langle\hat{\sigma}_{HV}(1)\rangle_w &= 0,\\
        \langle\hat{\sigma}_{HV}(2)\rangle_w &= 1
    \end{split}
\end{equation}
This is taken to mean that the polarisation travels on path 2, despite the photon travelling on path 1, in the postselected scenario,

Just from this initial description, we can see some issues with both the scenario and this interpretation: the polarisation discrimination operator considers polarisation in a different basis to the pre- and postselection; weak values are treated as directly providing information about system properties analogously to eigenvalues; and it is not obvious how we should best interpret the compound operators. We discuss these issues in Section \ref{sect:Meaning}.

As mentioned earlier in this section, this implementation is different to the implementation of the original Cheshire cat paper \cite{Aharonov2013Cheshire} in two ways. First, we removed a phase plate, which originally added a phase of $\pi/2$ on (the equivalent of) path 2 at the same time as the HWP---however, this phase is not necessary to observe the effect. Second, we flipped the direction of the protocol, so the preselected state is the entangled state ($\ket{E_{\mathrm{CC}}}$), and the postselected state is the Bell-local state ($\ket{D+}$). Given the time-symmetry of quantum mechanics, this has no effect on the protocol. Finally, as mentioned in the introduction, local measurement is important for contextual analysis---our goal here, by using contextuality theory, is to consider the relation between measurements, most of which should be local. Presenting the postselection as a simple unentangled measurement helps with this. The form we use is similar to that given in \cite{Vaidman2013Past,Aharonov2013Peculiar}.

\section{Contextual Analysis}\label{sect:Context}

The quantum Cheshire cat scenario involves considering polarisation in a different basis to that used in the pre- and postselection. The peculiar weak values obtained could be manifestations of this measurement incompatibility. This motivates us to consider the protocol from the perspective of contextuality, which provides a framework for linking paradoxical quantum effects with changes in measurement basis. We therefore consider how to represent the quantum Cheshire cat protocol as a contextuality argument, where there are a set of statements which are true individually about a photon in the protocol, but which together lead to a contradiction.

To do this, we start by working out how best to represent the system properties we infer from the weak values above.

\subsection{Individual Claims}

\begin{figure}[ht]
    \centering
    \includegraphics[width=0.75\linewidth]{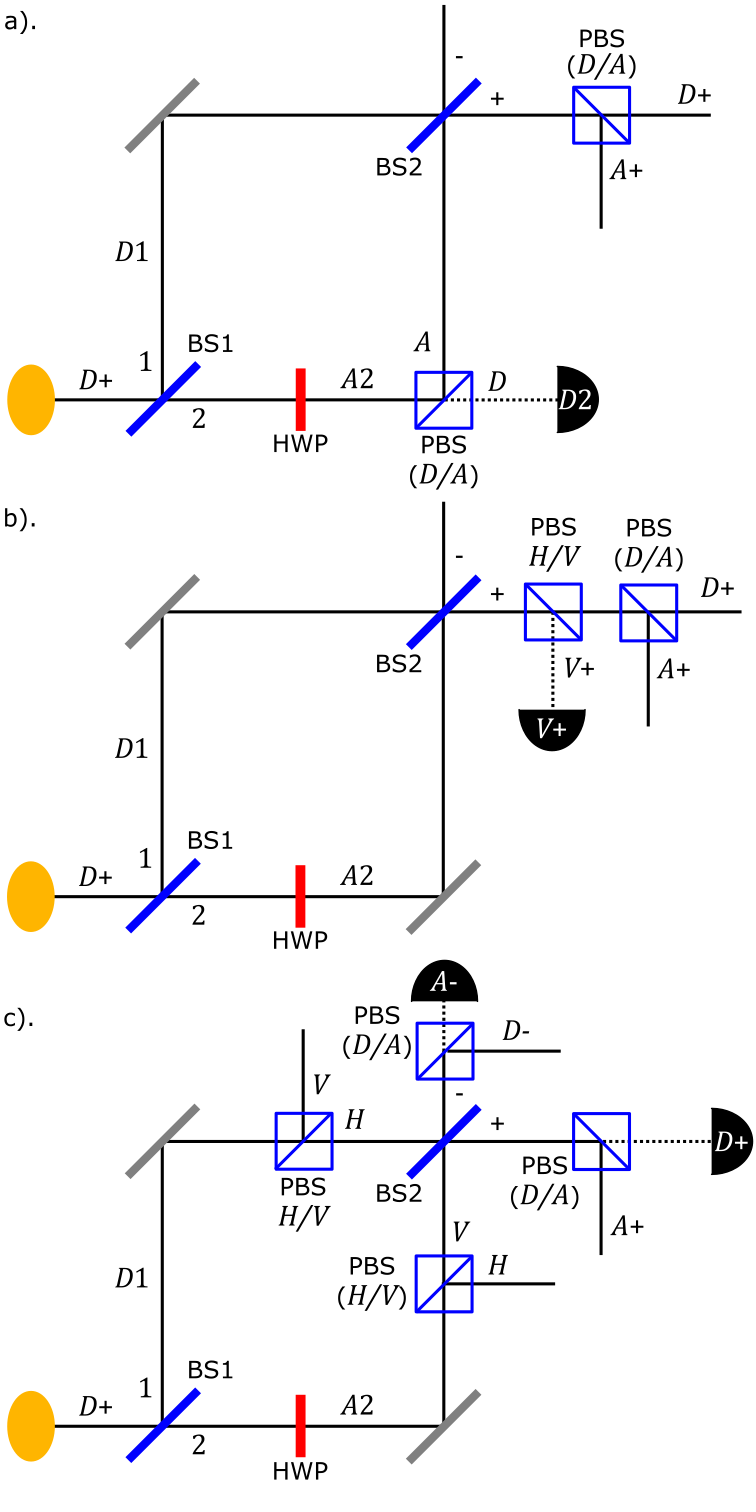}
    \caption{Altered forms of the quantum Cheshire cat protocol (as given in Fig.~\ref{fig:HolgerQCC}), each designed to allow the validation of one of the three claims made about the situation: Fig.~\ref{fig:Experiment}a allows us to test Claim \ref{Claim1}, Fig.~\ref{fig:Experiment}b allows us to test Claim \ref{Claim2}, and Fig.~\ref{fig:Experiment}c allows us to test Claim \ref{Claim3}. Given we are looking to show that preselected light never arrives at the relevant detectors, we can use either single photons or coherent states (e.g. laser light) to validate these claims. The black half-ovals are optical detectors, which based on their position will only detect light in the state given on their label (e.g. we know any light arriving at detector $A-$ in Fig.~\ref{fig:Experiment}c must be both $A$-polarised and in the ``$-$'' position superposition).}
    \label{fig:Experiment}
\end{figure}

The quantum Cheshire cat paradox can be represented by three key claims, each of which is the negation of the pre- and postselected particle having one of three properties:

\begin{claim}[NOT-\{2\}]\label{Claim1}
``No particle on Path 2.'' 
\end{claim}
This claim comes from observing that the postselection requires the photon to be $D$-polarised. The preselection forces path 1 to be $D$-polarised and path 2 $A$-polarised (and $D$ and $A$ are orthogonal), so we would expect that only light which has been on path 1 can go to detector $D+$. Therefore, a particle which passes both the pre- and postselection should not have been on path 2:
\begin{equation}
\begin{split}
    \braket{D2}{E_\text{CC}} &= 0,\\
    \braket{D+}{A2} &= 0
\end{split}
\end{equation}
We denote the property of the pre- and postselected particle having been on path 2 as \{2\}, and so call this claim NOT-\{2\}. (This was previously demonstrated by Saeed \emph{et al} {\cite{saeed2021quantum}}.)

The experimentally-verifiable condition for this claim is the absence of any $D$-polarised photons on path 2. A slight modification of the Cheshire cat protocol allows us to test this condition directly. As shown in Fig.~\ref{fig:Experiment}a, we can put a $D/A$ polarising beamsplitter on path 2, positioned so it removes any $D$-polarised light from path 2 and sends it to a detector. As $A$-polarised light wouldn't meet the postselection criterion, only light which goes to this detector would have gone along path 2 and still passed the postselection. However, the detector is expected to record no counts, confirming $\braket{D2}{E_\text{CC}}=0$. This procedure will not change the postselection probability $P(D+)$ since the observation of no photons in $D2$ does not change the state $\ket{E_\text{CC}}$.

The probability of a photon which meets the preselection being in state $D2$ (and so arriving at this new detector) is
\begin{equation}\label{eqD2Zero}
\begin{split}
    P(D2) = 0
\end{split}
\end{equation}

Claim \ref{Claim1} can be identified with the weak value of the projector on path 2 in the original Cheshire cat paper,
\begin{equation}\label{eqweakproj2}
\begin{split}
\langle\hat{\Pi}(2)\rangle_w=0
\end{split}
\end{equation}
Since we postselect on the polarisation $D$, we can replace the identity operator in the polarisation space with a projection operator on polarisation $D$,
\begin{equation}
\begin{split}
\langle\hat{\Pi}(2)\rangle_w=\langle\proj{D}\otimes\proj{2}\rangle_w
\end{split}
\end{equation}
Therefore, the probability of 0 in Eq.~(\ref{eqD2Zero}) ensures that the weak value in Eq.~(\ref{eqweakproj2}) is also zero.

The experiment in Fig.~\ref{fig:Experiment}a has some similarity to the one done by Denkmeyr et al \cite{Denkmayr2014QCCExp}. In that experiment, they put a blocker onto each of the two paths in turn, to work out which path neutrons took in their Cheshire cat interferometer. They inferred that the neutrons must travel along (their equivalent of) path 1, as only on that path did the presence of the blocker affect the detection intensity. Our set-up shows that this was possible due to the polarisation-path correlation---blocking path 2 merely blocks $A$-polarised light, whereas all of the $D$-polarised light is blocked when path 1 is blocked. The above analysis thus shows more details of the effect demonstrated in \cite{Denkmayr2014QCCExp}.

\begin{claim}[NOT-{\{$V$\}}]\label{Claim2}
    ``No $V$-polarised particle.''
\end{claim}
This claim comes from observing that 
\begin{equation}
    \begin{split}
        \ket{E_\text{CC}} =  \frac{1}{\sqrt{2}}\left(\ket{H+}+\ket{V-}\right)
    \end{split}
\end{equation}
This means the photon is only $V$-polarised if it is in path-superposition ``$-$''. As the postselected state $\ket{D+}$ is in the orthogonal path-superposition ``$+$'', a photon which passes both pre- and postselection cannot be $V$-polarised:
\begin{equation}
\begin{split}
    \braket{V+}{E_\text{CC}} &= 0,\\
    \braket{D+}{V-} &= 0
\end{split}
\end{equation}
We denote the property of the pre-and postselected particle having been $V$-polarised as \{$V$\}, and so call this claim NOT-\{$V$\}.

The experimentally-verifiable condition for this claim is the absence of any $V$-polarised photons in the ``$+$'' output of the interferometer. A slight modification of the Cheshire cat protocol allows us to test this condition directly. We can test this condition by putting a $H/V$ polarising beamsplitter just before the $D/A$ PBS, set up so any $V+$ light from the interferometer goes to a detector (see Fig.~\ref{fig:Experiment}b). Given any light in state $V-$ wouldn't meet the postselection condition, this ensures only light which goes to this detector could have been $V$-polarised and still passed the postselection. This procedure will not change the postselection probability $P(D+)$ since the observation of no photons in $V+$ does not change the state $\ket{E_\text{CC}}$.

The probability of a photon which meets the preselection being in state $V+$ (and so arriving at this new detector) is
\begin{equation}\label{eqV+Zero}
\begin{split}
    P(V+)= 0
\end{split}
\end{equation}

It may be worth noting that the original quantum Cheshire cat paper \cite{Aharonov2013Cheshire} does not make any statements about the polarisation itself. However, in close analogy to Claim \ref{Claim1}, we can now identify the weak value of the projector on $V$-polarisation,
\begin{equation}\label{eq.Vweak}
\begin{split}
\langle\hat{\Pi}(V)\rangle_w=0
\end{split}
\end{equation}
where $\hat{\Pi}(V) = \proj{V}\otimes\mathds{1}$.
As before, the postselection of $+$ ensures this weak value is the same as the weak value of $V+$, which is zero because there is no $V+$ component in $\ket{E_\text{CC}}$.

\begin{claim}[NOT-{\{$\Phi$\}}]\label{Claim3}
    ``No $\Phi$-correlation of path and polarisation.''
\end{claim}

This Claim needs to establish the relation between polarisation and path. To classify the correlations in a way that allows us to connect the $H/V$ and path basis to the postselected Bell states, we consider the Bell states. The two Bell states for which path 1 is always $H$-polarised and path 2 is always $V$ polarised are states $\Phi^+$ and $\Phi^-$, where
\begin{equation}
    \begin{split}
        \ket{\Phi^\pm} = \frac{1}{\sqrt{2}}\left(\ket{H1}\pm\ket{V2}\right)
    \end{split}
\end{equation}
We therefore refer to this correlation between path and polarisation as $\Phi$-correlation.

It is now possible to show there cannot be any $\Phi$-correlation, as the preselection of $\ket{E_\text{CC}}$ does not include $\ket{\Phi^+}$, and the postselection of $\ket{D+}$ does not include any $\ket{\Phi^-}$:
\begin{equation}
\begin{split}
    \braket{\Phi^+}{E_\text{CC}} &= 0,\\
    \braket{D+}{\Phi^-} &= 0
\end{split}
\end{equation}
This relation shows that the postselected outcome $\ket{D+}$ does not contain any $\ket{\Phi^-}$ component. This means that the absence of the $\ket{\Phi^+}$ component in the preselected initial state $\ket{E_\text{CC}}$ is sufficient to to experimentally confirm Claim \ref{Claim3}.

We denote the property of the pre- and postselected particle having the $\Phi$-correlation between polarisation and path as \{$\Phi$\}, and so call this claim NOT-\{$\Phi$\}.

The experimentally-verifiable condition for this claim is the absence of $\ket{\Phi^+}$. Since this is an entangled state, it is best to verify it in a two-step process. As shown in Fig.~\ref{fig:Experiment}c, we can first remove all components that are not $\Phi$-correlated, by using the appropriate PBSs to remove $V1$ and $H2$. This corresponds to a projection operator $\hat{\Pi}(\Phi)$ to the input state, where
\begin{equation}
    \hat{\Pi}(\Phi) = \proj{H1}+\proj{V2}
\end{equation}
Since the state $\ket{\Phi^-}$ does not include any $D+$ component, a detection of $D+$ in the output now corresponds to a detection of $\Phi^+$ in the initial state:
\begin{equation}
    \bra{D+}\hat{\Pi}(\Phi) = \frac{1}{\sqrt{2}}\bra{\Phi^+}
\end{equation}
The probability of the photon which meets the preselection being in state $\Phi^+$ is
\begin{equation}\label{eqPhi+Zero}
\begin{split}
    P(\Phi^+)= 0
\end{split}
\end{equation}
This shows that the part of the input state $\ket{E_\text{CC}}$ that has a $\Phi$-correlation does not connect to the postselected outcome $D+$.

It is also possible to identify this claim with the weak value of the projector $\hat{\Pi}(\Phi)$,
\begin{equation}\label{eq:weakphizero}
    \langle\hat{\Pi}(\Phi)\rangle_w = 0
\end{equation}
where the postselection of $D+$ ensures that this is the same as the weak value of the projector on $\Phi^+$, which is zero because there is no $\Phi^+$ component in $\ket{E_\text{CC}}$.

\subsection{Combining Claims}

\begin{figure}[ht]
    \centering
    \includegraphics[width=\linewidth]{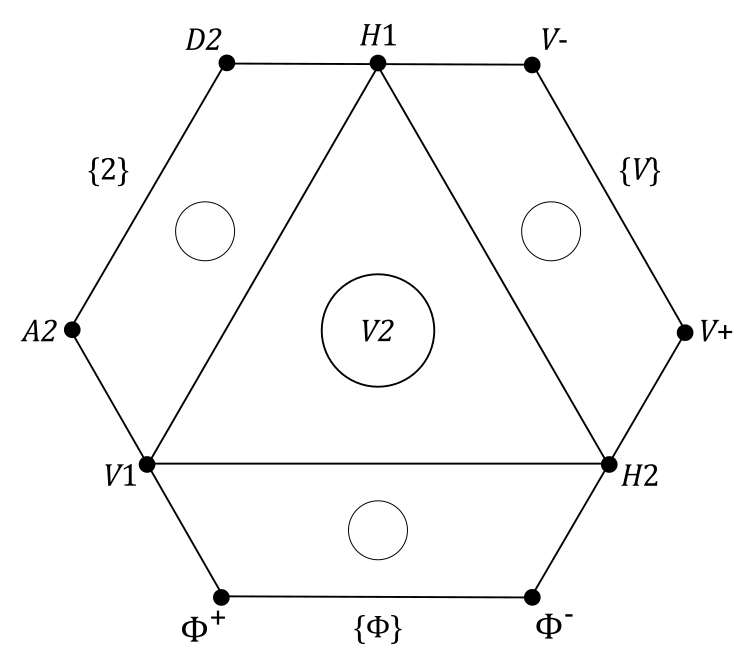}
    \caption{Hexagram diagram of the contextuality relations for the quantum Cheshire cat scenario. Contexts are represented by minimal closed cycles (with circles within these for clarity)---all states in the same context are all mutually orthogonal. Each of the three outer contexts is comprised of four states: two which share a side of the inner triangle, and two which share a side of the hexagon. The three outer contexts each correspond to one of the three Claims, where the prohibited properties are indicated by \{2\}, \{$V$\}, and \{$\Phi$\}, placed between the two states in the context which share this property. All of the states that satisfy one of the claims are part of the inner triangle---specifically, each state is uniquely defined by two of the Claims. There is no state for which all three Claims are true, since each state of the inner triangle is orthogonal to the two states that satisfy the claim opposite to it.}
    \label{fig:QCCHex}
\end{figure}

The typical structure of a contextuality argument involves identifying claims which apply individually to a situation, even though their combination would result in logical contradictions. We do this here for the quantum Cheshire cat scenario by showing that the three claims discussed above cannot be satisfied by any combination of $H/V$-polarisation in path 1 or 2.

By combining the claims we define above in certain ways, we see that each pair of claims uniquely infers a different state:

\begin{inference}[$H1$]\label{Inf1}
    Claim \ref{Claim1} + Claim \ref{Claim2} $\rightarrow$ $H1$
\end{inference}
This inference comes from Claim \ref{Claim1} saying the particle wasn't on path 2 (so must be on path 1), and Claim \ref{Claim2} saying the particle wasn't $V$-polarised, so it must be $H$-polarised. $H1$ is the only state which agrees with these two claims.
    
 \begin{inference}[$H2$]\label{Inf2}
    Claim \ref{Claim2} + Claim \ref{Claim3} $\rightarrow$ $H2$
\end{inference}
This inference comes in multiple steps. Claim \ref{Claim3} tells us the particle wasn't in a $\Phi$-correlation, so wasn't in either state $H1$ or $V2$. Therefore, the particle must have been in either state $H2$ or $V1$. Claim \ref{Claim2} then adds that the particle wasn't $V$-polarised, so of those two states, it can't have been in state $V1$. Therefore, the particle must have been in state $H2$.
    
\begin{inference}[$V1$]\label{Inf3}
    Claim \ref{Claim3} + Claim \ref{Claim1} $\rightarrow$ $V1$
\end{inference}
Again, this inference comes in multiple steps. Claim \ref{Claim3} says the particle wasn't in a $\Phi$-correlation, so it wasn't in either state $H1$ or $V2$. Therefore, the particle must have been in either state $H2$ or $V1$. Claim \ref{Claim1} then adds that the particle wasn't on path 2, so of those two states, it can't have been in state $H2$. Therefore, the particle must have been in state $V1$.

These inferences all contradict one another---Inference \ref{Inf1} infers the state is $H1$, Inference \ref{Inf2} infers the state is $H2$, and Inference \ref{Inf3} infers the state is $V1$. We can represent this as a contextuality diagram, as shown in Fig.~\ref{fig:QCCHex}. This diagram shows that the quantum Cheshire cat can be represented as a typical example of contextuality. It is interesting to compare this to the original formulation, which lacks the symmetry of the contextuality diagram. This broken symmetry results from the combination of Claims in Inference \ref{Inf2}. As we argued in the previous subsection, Claim \ref{Claim1} corresponds to the weak values of the spatial projection operator in the original quantum Cheshire cat paper \cite{Aharonov2013Cheshire}---specifically, since the weak values of projections of path 1 and path 2 must add up to 1, the weak value of 0 for projection on path 2 necessarily corresponds to a weak value of 1 for projection on path 1. To combine the weak values given in Claims \ref{Claim2} and \ref{Claim3}, note that each of the projectors can be separated into two parts. According to Eq.~(\ref{eq.Vweak}) in Claim \ref{Claim2},
\begin{equation}\label{eq:weakprojv1v2}
    \langle\proj{V1}\rangle_w = -\langle\proj{V2}\rangle_w
\end{equation}
Similarly, Eq.~(\ref{eq:weakphizero}) in Claim \ref{Claim3} gives
\begin{equation}\label{eq:weakprojh1v2}
    \langle\proj{H1}\rangle_w = -\langle\proj{V2}\rangle_w
\end{equation}
The weak value which is supposed to indicate the absence of polarisation in path 1 is given by
\begin{equation}\label{eq:weakHVproj1}
    \langle\hat{\sigma}_{HV}\otimes\proj{1}\rangle_w = \langle\proj{H1}\rangle_w -  \langle\proj{V1}\rangle_w
\end{equation}
Eqs.~(\ref{eq:weakprojv1v2}) and (\ref{eq:weakHVproj1}) show the two terms on the right hand side of Eq.~({\ref{eq:weakHVproj1}}) are equal, so that they cancel, and the weak value in Eq.~(\ref{eq:weakHVproj1}) is indeed zero.
It follows from Eq.~(\ref{eq.Vweak}) in Claim \ref{Claim2} that the weak value of the projector on $V$-polarisation is 0, and so the weak value of the projector on $H$-polarisation is 1. The weak value of $\hat{\sigma}_{HV}$ is therefore 1, and so
\begin{equation}
    \langle\hat{\sigma}_{HV}\otimes\proj{1}\rangle_w + \langle\hat{\sigma}_{HV}\otimes\proj{2}\rangle_w = 1
\end{equation}
This means
\begin{equation}
    \langle\hat{\sigma}_{HV}\otimes\proj{2}\rangle_w = 1
\end{equation}
as in the original quantum Cheshire cat paper \cite{Aharonov2013Cheshire}.

\subsection{Experimental Verification of Contextuality}

Fig.~\ref{fig:Experiment} suggests that all of the Claims can be verified experimentally. However, in a realistic experimental situation, the probabilities of the prohibited outcomes will not be exactly zero. It is therefore useful to formulate quantum contextuality as an inequality violation \cite{Simon2001Inequality,Larsson2002Ineq,Cabello2008Ineq,Klyachko2008Ineq,ji2023characterization,Matsuyama2023Ineq}. The postselected outcome $D+$ appears to be impossible when all three claims are satisfied. Statistically, the probability of finding $D+$ should therefore be upper-bounded by the sum of the probabilities that each of the Claims is violated,
\begin{equation}\label{Eq.NonContextIneq}
P(D+) \leq P(V+)+P(D2)+P(\Phi^+).
\end{equation}
Fig.~\ref{fig:Experiment} shows the experimental methods for determining the individual probabilities. In the ideal case, the probability of finding the outcome $D+$ in the input state $\ket{E_\text{CC}}$ should be 1/4, where all of the probabilities on the right side of the inequality will be close to zero.

\section{Coherences between prohibited states}\label{sect:Coherences}

The original interpretation of the quantum Cheshire cat paradox is that the polarisation becomes disembodied from the particle. However, the contextuality analysis above indicates that it may be problematic to consider path 2 to be empty, just as it would be problematic to consider the particle to be entirely $H$-polarised. Let us consider whether the weak values formalism indicates why and how path 2 and $V$-polarisation are involved in this paradoxical behaviour.
To do so, we note that all of the claims eliminate a possibility by combining probabilities of zero in the preselected state with probabilities of zero in the postselected state. This results in weak values of zero, because for any statement represented by a projector $\hat{\Pi}$, the weak value will be zero when either the initial or the final state are orthogonal to a set of eigenstates of the projector.

The weak value in Claim \ref{Claim2} can be separated into the weak values of projectors on $A2$ and $D2$. The weak value of the projector on $A2$ is zero because it is orthogonal to the final state; the weak value of $D2$ is zero because it is orthogonal to the initial state. However, we cannot use this observation to determine the weak values of projectors on $H2$ and $V2$, despite the two also summing to zero. This is because both $V2$ and $H2$ have non-zero components in the initial and the final state. Weak values are determined by combining these non-zero components in a product. The weak values of zero for the projectors of $A2$ and $D2$ correspond to a contribution from ordered coherences between the two.

Making use of the orthogonality of $A2$ and the postselected state $D+$, and the orthogonality of $D2$ and the initial state $\ket{E_\text{CC}}$, the weak values of the projectors on $H2$ and $V2$ can be represented as weak values of coherences between $A2$ and $D2$, 
\begin{equation}
\begin{split}
    \langle\proj{H2}\rangle_w &= \frac{1}{2}\langle\ketbra{D2}{A2}\rangle_w,\\
    \langle\proj{V2}\rangle_w &= -\frac{1}{2}\langle\ketbra{D2}{A2}\rangle_w
     \end{split}
\end{equation}
Here, the weak value of an operator (be it a projector or a coherence) can be described by
\begin{equation}
    \langle\hat{O}\rangle_w = \text{Tr}\left(\hat{O}\frac{\ketbra{i}{f}}{\braket{f}{i}}\right)
\end{equation}
This shows that weak values can be represented as expectation values of a statistical operator, representing the pre- and postselection \cite{Hofmann2010PostselStats}. This statistical operator can be decomposed using the weak values of the coherences between the states $n$ and $m$ of a given basis \cite{Hofmann2011ComplexPhases,Hofmann2012ComplexJointProbs}:
\begin{equation}
\begin{split}
    \forall n,m\;\text{such that}\;|\braket{n}{m}|^2 = \delta_{n,m},\\
    \frac{\ketbra{i}{f}}{\braket{f}{i}} = \sum_{n,m}\langle\ketbra{m}{n}\rangle_w\ketbra{n}{m}
\end{split}
\end{equation}

For the quantum Cheshire cat scenario, we can decompose the statistical operator in different orthogonal bases, corresponding to the different Claims. If we decompose it in the $(D/A)\otimes(1/2)$ basis, we get
\begin{equation}
    \begin{split}
        &\frac{\ketbra{E_\text{CC}}{D+}}{\braket{D+}{E_\text{CC}}}\\
        &=\proj{D1}+\ketbra{D1}{A2}+\ketbra{D2}{D1}+\ketbra{D2}{A2}\\
    \end{split}
\end{equation}
Given this decomposition only has a projector onto a state on path 1, this appears to support Claim \ref{Claim1}. However, the decomposition also has a coherence between two states which are both on path 2.

Similarly, if we decompose the statistical operator in the $(H/V)\otimes(+/-)$ basis, we get
\begin{equation}
    \begin{split}
        &\frac{\ketbra{E_\text{CC}}{D+}}{\braket{D+}{E_\text{CC}}}\\
        &= \proj{H+}+\ketbra{H+}{V-}+\ketbra{V+}{H+}\\
        &\;\;\;+\ketbra{V+}{V-}
    \end{split}
\end{equation}
Given this decomposition only has a projector onto the $H$-polarisation, this appears to support Claim \ref{Claim2}. However, the decomposition also has a coherence between two states which are both $V$-polarised.

Finally, if we decompose it in the Bell-basis, we get:
\begin{equation}
\begin{split}
    &\frac{\ketbra{E_\text{CC}}{D+}}{\braket{D+}{E_\text{CC}}}\\
    &= \proj{\Psi^+}+\ketbra{\Phi^+}{\Psi^+}+\ketbra{\Psi^+}{\Phi^-}+\ketbra{\Phi^+}{\Phi^-}
\end{split}
\end{equation}
Given this decomposition only has a projector onto the Bell state $\ket{\Psi^+}$, this appears to support Claim \ref{Claim3}. However, the decomposition also has a coherence between two states which are both $\Phi$-correlated.

These three decompositions show that the statistical operator contains coherences between prohibited states (specifically between $A2$ and $D2$ in the $(D/A)\otimes(1/2)$ decomposition, between $V-$ and $V+$ in the $(H/V)\otimes(+/-)$ decomposition, and between $\Phi^-$ and $\Phi^+$ in the Bell-basis decomposition). These coherences are between states which are prohibited---specifically, between one state prohibited by the postselection, and another prohibited by the preselection. By having the state prohibited by preselection as its ``ket'', and the state prohibited by postselection as its ``bra'', the coherence is allowed by both the pre- and postselection, despite the two states it is formed of both individually being prohibited. As discussed in \cite{hofmann2023quantum}, these coherences between prohibited states form necessary components of the quantum descriptions of contextual systems, despite not being describable classically.

By changing basis, we see that all three of these coherences contain the projector $\proj{V2}$, which has an anomalous weak value of -1/2 (as discussed in \cite{Hofmann2015Paradoxes}). Given the weak values of $\proj{H2}$, $\proj{V1}$, and $\proj{H1}$ are all +1/2, this negative weak value seems to cancel out the projection onto each of these states, and so prohibit the photon from having the property shared by $V2$ and that state (\{2\} for $H2$, \{$V$\} for $V1$, and \{$\Phi$\} for $H1$).

However, each of these coherences only exists in one measurement basis: if we measure in the $(D/A)\otimes(1/2)$ basis, $\proj{V2}$ only cancels out $\proj{H2}$, leaving the coherence $\ketbra{D2}{A2}$; if we measure in the $(H/V)\otimes(+/-)$ basis, $\proj{V2}$ only cancels out $\proj{V1}$, leaving the coherence $\ketbra{V+}{V-}$; and if we measure in the Bell-basis, $\proj{V2}$ only cancels out $\proj{H1}$, leaving the coherence $\ketbra{\Phi^+}{\Phi^-}$. There is only one -1/2 term, but three +1/2 terms---but, in each basis, one of those +1/2 terms is cancelled out, and so it appears that the shared property of $\proj{V2}$ with that state is prohibited.

This is different from a noncontextual prohibition of a property, where any state with that property should have no projector for \emph{any} choice of basis, not just for one choice of basis. This shows why coherences are important---the projector for a state with a given property in a contextual scenario can hide in the coherence; suppressed in that choice of basis (and so for that measurement, or question), but stopping us from holding the negation of both that property and another contextual property true simultaneously.

A single negative-projector state can only cancel one other state at a time, when, for it to be non-contextually true that the scenario doesn't possess these properties, all three would need to be cancelled. These coherences are therefore responsible for the mysterious violation of the noncontextual inequality (Eq.~(\ref{Eq.NonContextIneq})), by stopping us from being able to ensure $P(D2)$, $P(V+)$ and $P(\Phi^+)$ are all zero simultaneously: forcing one of these states to be probability zero frees the non-zero projectors of the other two states, and so allows the other two probabilities to be non-zero.

The negative weak value of state $V2$ implies the photon being in this state is suppressed more heavily than is classically possible, to the extent that it suppresses the photon being $V$-polarised (by cancelling out the positive weak value of $V1$), or being on path 2 (by cancelling out the positive weak value of $H2$), or having $H1/V2$ correlation (by cancelling out the positive weak value of $H1$), depending on how the photon is measured. The weak value of $V2$ being negative therefore indicates the system's behaviour will be contextual. The suppression of each of the properties of the photon being in state $V2$ (the photon being $V$-polarised, the photon being on path 2, and the photon having a $H1/V2$ correlation), relates to one of three contexts. Each context is linked to both of the other contexts (they each share one state with each other contexts); however, no state is in all three contexts. Therefore, the photon is never in a state where all three of these properties are suppressed simultaneously. This means things we would classically infer from all three of these properties being suppressed (such as the probability of a photon that passes the preselection being in state $D+$ being lower than the sum of the probabilities of it being in states $D2$, $V+$, and $\Phi^+$), can be shown not to be the case (as per the violation of Eq.~(\ref{Eq.NonContextIneq})).

\section{Compound Operators and Basis Bias}\label{sect:Meaning}

A major result of this paper is showing that we can talk about the quantum Cheshire cat scenario as a contextual scenario. However, being able to do so relies on being able to characterise the polarisation/path correlation prohibited by the compound operators' weak values. The quantum Cheshire cat is hard to understand precisely because this condition---the prohibition of the $\Phi$-correlation---is hard to extract from the original presentation of the scenario.

The main evidence for a paradox given in the original quantum Cheshire cat paper \cite{Aharonov2013Cheshire} is from weak values---there is no way to directly measure the polarisation or path-presence without disturbing the system, so we can observe no direct evidence of this paradox; and as one of us has discussed previously, inferring from weak values to properties is non-trivial at best \cite{Hance2023Weak}. However, given the weak value-related issues with the quantum Cheshire cat have been discussed heavily (see Duprey et al's review \cite{duprey2018quantum}), and often rely on an understanding of weak values as requiring weak measurement to obtain (which we now know not to be the case \cite{hofmann2012weak,hofmann2014sequential,Cohen2018WeakStrong,wagner2023quantum}), we will focus on the problem of the construction of the compound operator, which reduces the Cheshire cat problem to only two weak values.

The claim that the photon separates from its polarisation in the quantum Cheshire cat scenario relies on a lack of symmetry. There is no mathematical difference between the path-presence qubit and the polarisation qubit in the scenario---both are just qubits. However, we usually interpret a photon's position very differently from its polarisation. This is expressed by the compound operators given in Eq.~(\ref{eq:Compound}). We can now relate these compound operators to the weak values of projectors,
\begin{equation}
    \begin{split}
        \langle\proj{H1}\rangle_w &= +1/2,\\
        \langle\proj{H2}\rangle_w &= +1/2,\\
        \langle\proj{V1}\rangle_w &= +1/2,\\
        \langle\proj{V2}\rangle_w &= -1/2
    \end{split}
\end{equation}
The weak value of the compound operator intended to describe the presence of polarisation in path 2 is then given by
\begin{equation}
    \begin{split}
        \langle\hat{\sigma}_{HV}\otimes\proj{2}\rangle_w = \langle\proj{H2}\rangle_w - \langle\proj{V2}\rangle_w
    \end{split}
\end{equation}
We can now see that the presence of a polarisation in path 2 is a direct consequence of the negative weak value of the projector on $V2$. The apparent contradiction with Claim \ref{Claim1} is expressed by the weak value of the projector on path 2, which can be written as
\begin{equation}
    \langle\mathds{1}\otimes\proj{2}\rangle_w = \langle\proj{H2}\rangle_w+\langle\proj{V2}\rangle_w
\end{equation}
The contradiction between these two equations is a direct result of the coherence between $A2$ and $D2$ found when describing the pre- and postselection in the $(D/A)\otimes(1/2)$ basis. This specific basis then expresses a combination of Claim \ref{Claim1} with Inference \ref{Inf2}. Other combinations are possible and can be constructed by basing the paradox on a different initial Claim. If we base the paradox on Claim \ref{Claim2}, the compound operator is
\begin{equation}
    \langle\proj{V}\otimes\hat{\sigma}_\pm\rangle_w = \langle\proj{V1}\rangle_w - \langle\proj{V2}\rangle_w
\end{equation}
where $\hat{\sigma}_\pm$ is path difference operator
\begin{equation}
\begin{split}
    \hat{\sigma}_{\pm} &= \proj{+}-\proj{-}\\
    &= \ketbra{1}{2}+\ketbra{2}{1}
\end{split}
\end{equation}
The contradiction is that we find a path-bias in the polarisation $V$, even though according to Claim \ref{Claim2} there is no $V$-polarisation, as given by the sum of the weak values
\begin{equation}
    \langle\proj{V}\otimes\mathds{1}\rangle_w = \langle\proj{V1}\rangle_w + \langle\proj{V2}\rangle_w
\end{equation}
This version of the paradox can be traced back to the weak value of the coherence between $V+$ and $V-$. We therefore see that changing the Claim on which we base the paradox is equivalent to changing the basis.

If we base the paradox on Claim \ref{Claim3}, the compound operator describes a bias between $H1$ and $V2$ \emph{within} the $\Phi$-correlation:
\begin{equation}
    \langle\hat{B}_\Phi\rangle_w = \langle\proj{H1}\rangle_w-\langle\proj{V2}\rangle_w
\end{equation}
The contradiction is that we find such a bias in the $\Phi$ correlation, even though there seems to be no $\Phi$-correlation present, as given by the sum of the weak values,
\begin{equation}
    \langle\hat{\Pi}(\Phi)\rangle_w = \langle\proj{H1}\rangle_w+\langle\proj{V2}\rangle_w
\end{equation}
This version of the paradox can be traced back to the weak value of the coherence between $\Phi^+$ and $\Phi^-$. It may be worth noting that the basis here is a basis of entangled states.

The analysis above shows that the original quantum Cheshire cat paradox is based on an arbitrary choice of a conditional bias, which combines two of the three Claims. The main reason why the initial choice was Claim \ref{Claim1} is that it corresponds to the basis choice of the 1/2 basis, which expresses our intuitive bias towards the idea that particles are naturally localised.

The formulation of all three versions of the paradox involves the contradiction between the weak value of a projector, and the weak value of a compound operator corresponding to a conditional bias. In all three cases, the conditional bias is away from the state $V2$.

What does this mean? 
In the original quantum Cheshire cat paper \cite{Aharonov2013Cheshire}, the weak value of the compound operators supposedly tell us that the ``polarisation'' is disembodied onto path 2. It seems harder to claim that when we base the paradox on Claim \ref{Claim2}, the ``localisation'' of the photon is disembodied into $V$-polarisation.

Clearly, the problem is that we find it easier to imagine that empty space is somehow polarised than to imagine that an ``empty'' polarisation could be distributed in space. However, both concepts are actually equivalent. It is equally wrong to identify a particle with its position as it is to identify a particle with its polarisation. As a result, it is even possible to identify the correlation between position and polarisation as the fundamental property on which we can base the paradox. Ultimately, the paradox originates from a contradiction between these three features of the particle, as shown in full by the contextuality scenario involving all three claims.

\vspace{6pt}

\section{Conclusion}\label{sect:Conc}
We have shown that the quantum Cheshire cat scenario is contextual. We did this by introducing three separate Claims regarding the path and polarisation of a photon within the quantum Cheshire cat protocol. Each of these Claims can be tested experimentally, and the quantum Cheshire cat paradox can be verified without any recourse to weak values. We also showed that the paradox can be expressed as a contradiction between only two statements, by combining any two claims using inference, where the disembodied polarisation corresponds to a combination of Claim \ref{Claim1} with Inference \ref{Inf2}. The original quantum Cheshire cat paradox can thus be recovered from the general formulation we present.

We then went on to consider \emph{why} the scenario is contextual. By decomposing a statistical operator, formed by the pre- and postselection, in different bases corresponding to different measurement contexts, we observed coherences between states that were prohibited by either the pre- or postselection. We showed that these coherences between prohibited states cause the contextual behaviour---each of these coherences can be transformed into a projector on some state (depending on the context), minus a projector on state $V2$. This minus, corresponding to the negative weak value of $V2$, allows it to ``cancel'' out that state, but only in that context. Each of the three relevant contexts can be identified with one state that is cancelled out by a negative weak value contribution from $V2$. The measurement performed on the system determines which cancellation is relevant. Therefore, the system not having the property shared by that state and state $V2$ only occurs when the system is measured in that context.

In this paper, we have clarified how the quantum Cheshire cat paradox should be interpreted---specifically that the argument that the polarisation becomes ``disembodied'' results from only considering one specific pairing of the three mutually-incompatible properties in what is ultimately just a contextual system. This analysis allows us to properly relate the quantum Cheshire cat paradox to fundamental properties of quantum mechanics, including contextuality, weak values, and coherences between prohibited states. Investigating these relations further presents interesting new directions for research in quantum foundations.

\vspace{6pt}

\textit{Acknowledgements -} We thank Masataka Iinuma and Tomonori Matsushita for helpful comments on early versions of this paper. JRH thanks Prof James Ladyman and Prof John Rarity for useful earlier discussions of the quantum Cheshire cat paradox. JRH acknowledges support from Hiroshima University's Phoenix Postdoctoral Fellowship for Research, the University of York's EPSRC DTP grant EP/R513386/1, and the Quantum Communications Hub funded by EPSRC grants EP/M013472/1 and EP/T001011/1. MJ acknowledges support from JST SPRING, Grant Number JPMJSP2132. 

\bibliographystyle{unsrturl}
\bibliography{ref.bib}

\end{document}